\documentclass[12pt]{article}
\begin{document}
\newfont{\elevenmib}{cmmib10 scaled\magstep1}%
\renewcommand{\theequation}{\arabic{section}.\arabic{equation}}

\newcommand{\preprint}{
            \begin{flushleft}
   \elevenmib Yukawa\, Institute\, Kyoto\\
            \end{flushleft}\vspace{-1.3cm}
            \begin{flushright}\normalsize  \sf
            YITP-01-60\\
            KUCP-0191\\
   {\tt hep-th/0109008} \\ September 2001
            \end{flushright}}
\newcommand{\Title}[1]{{\baselineskip=26pt \begin{center}
            \Large   \bf #1 \\ \ \\ \end{center}}}
\newcommand{\Author}{\begin{center}\large \bf
             R.\, Sasaki$^a$
     and K.\, Takasaki$^b$\end{center}}
\newcommand{\Address}{\begin{center} \it
            $^a$ Yukawa Institute for Theoretical Physics, Kyoto
            University,\\ Kyoto 606-8502, Japan \\
     $^b$  Department of Fundamental Sciences,
            Faculty of Integrated Human Studies,\\ Kyoto University,
            Kyoto 606-8501,Japan
      \end{center}}
\newcommand{\Accepted}[1]{\begin{center}{\large \sf #1}\\
            \vspace{1mm}{\small \sf Accepted for Publication}
            \end{center}}
\baselineskip=20pt

\preprint
\thispagestyle{empty}
\bigskip
\bigskip
\bigskip
\Title{Quantum Inozemtsev model, quasi-exact solvability and ${\cal
N}$-fold supersymmetry}
\Author

\Address
\vspace{2cm}

\begin{abstract}
Inozemtsev models are classically integrable multi-particle dynamical
systems
related to Calogero-Moser models.
Because of the additional $q^6$ (rational models) or $\sin^22q$
(trigonometric
models) potentials, their quantum versions are not exactly solvable in
contrast with
Calogero-Moser models. We show that quantum Inozemtsev models
can be deformed to be
a widest class of partly solvable (or quasi-exactly solvable) multi-particle
dynamical systems.
They posses ${\mathcal N}$-fold supersymmetry
which is equivalent to quasi-exact
solvability. A new method for identifying and solving
quasi-exactly solvable systems,
the method of pre-superpotential,
is presented.
\end{abstract}

\newpage
\section{Introduction}
\label{intro}
In this paper we address the problem of the relationship
among classical integrability, quantum integrability and
quantum partial integrability (or quasi-exact solvability)
within the multi-particle dynamical systems
by taking a wide class of explicit examples, Inozemtsev models.
We demonstrate that  the Inozemtsev models (with degenerate potentials,
that is non-elliptic potentials) can be made quantum partly solvable
(or QES, quasi-exactly solvable).
For this purpose we present a simple new formulation of QES systems
of single as well as multiple degrees of freedom.
We also show that the notions of higher derivative
(or non-linear or ${\mathcal N}$-fold) supersymmetry and
quasi-exact solvability are equivalent. In other
words, Inozemtsev models provide plentiful examples of multi-particle
${\mathcal N}$-fold supersymmetry.

Inozemtsev models \cite{bib:ad77}-\cite{Ino0} are generalisation of
Calogero-Moser models
\cite{Cal,Sut,CalMo},\cite{OP1,OPq,bcs2},\cite{DHoker_Phong,bcs1}
associated with the root system of $BC$ type and $A$ type.
They belong to the category of `twisted' Calogero-Moser models
\cite{DHoker_Phong, bst}.
Like all the Calogero-Moser models they are classically integrable
for all the four types of potentials, elliptic, hyperbolic, trigonometric
and
rational in the sense that their equations of motion can be expressed
in Lax pair forms.

  We start by asking
a general and naturally vague question: to which extent does classical
integrability imply quantum integrability in multi-particle dynamical
systems?
As is well-known, the converse, that  quantum integrability always implies
classical integrability, is true in multi-particle dynamical systems, since
the
quantum system is the $\hbar$ deformation of the classical one.
We do not know a way to answer this question abstractly by starting from the
pure notion of classical integrability, although
some attempts have been made to construct quantum conserved quantities
as a deformation of classical ones in the
framework of perturbed conformal field
theory (see, for example, \cite{sasya}).

However, the experience of Calogero-Moser models,
the widest class of integrable
multi-particle systems ever known, tells that the classical integrability
is very close to quantum integrability.
The quantum integrability is proved universally, that is for all the root
systems
including the non-crystallographic ones, for Calogero-Moser models
with degenerate potentials \cite{kps,HeOp}, namely those with trigonometric,
hyperbolic and rational potentials.

Thus we are naturally led to the question of quantum integrability of
Inozemtsev
models with degenerate potentials.
Rational Inozemtsev models have an additional potential of sixth degree
polynomial in $q$ (coordinates), see (\ref{eq:exham2}), on top of the
Calogero-Moser potentials.
Trigonometric Inozemtsev models have an additional $\sin^22q$ potential, see
(\ref{eq:exham3}), on top of the
Calogero-Moser potentials. These additional potentials destroy the
mechanism for providing quantum conserved quantities,
the so-called `sum-to-zero'
condition of the second member of the Lax pair \cite{bms,kps}.

The very interactions ($q^6$, $\sin^22q$, etc) that constitute obstructions
for
quantum integrability of Inozemtsev models are known to play essential
roles in the quasi-exactly solvable \cite{turush} one particle quantum
mechanics. This leads to a conjecture that {\em at least a certain class of
Inozemtsev models can be made quasi-exactly solvable\/}. We
demonstrate that {\em supersymmetrisable} (see (\ref{susyla1}) and
(\ref{susyla2})) Inozemtsev models can be deformed to quasi-exactly solvable
systems which are characterised by an integer deformation
parameter ${\mathcal M}$. It is also shown that the concepts of quasi-exact
solvability and {\em higher derivative\/}
\cite{ais} or {\em non-linear\/}
\cite{plyu} or
${\mathcal N}$-{\em fold} \cite{aoy} supersymmetry (with a typical
relationship ${\mathcal N}={\mathcal M}+1$) are equivalent.

\bigskip
This paper is organised as follows.
In section two we present the basic tool for investigating quasi-exactly
solvable systems which we call the {\em method of pre-superpotential\/}.
We briefly summarise the classical Inozemtsev models
in comparison with Calogero-Moser models. In
section three we demonstrate the quasi-exact solvability of a single
particle rational
$BC$ type Inozemtsev model based on a new method of employing
a pre-superpotential
$W$. This provides the most general single particle QES system with
$q^6$ potential known to
date.
The equivalence of quasi-exact solvability and
${\mathcal N}$-fold supersymmetry
is generally established. Other related notions, the ``Bethe Ansatz" type
equations \cite{muw}, ODE spectral equivalence \cite{ddt,jsuz} and
Bender-Dunne polynomials \cite{bender} are simply explained from the new
point of view. Section four deals with the quasi-exact solvability of
a single particle trigonometric
$BC$ type Inozemtsev model. In section five  we discuss the rational $A$
type
Inozemtsev model with $q^4$ potential, which provides an example of a
spontaneously broken ${\mathcal N}$-fold supersymmetry.
Through section six to section eight, various Inozemtsev models (rational
$BC$
type, trigonometric $BC$ type and trigonometric $A$ type) are shown to be
quasi-exactly solvable and the generators of
${\mathcal N}$-fold supersymmetries
are identified.
The QES of quantum Inozemtsev models is the consequence of exact solvability
of quantum Calogero-Moser models and QES of the added single particle type
interactions. The final section is devoted for comments and discussion.
Appendix A gives the classical Lax pairs for the $BC$ type and $A$ type
Inozemtsev models in the same notation as used in the main text.
Appendix B is for the details of the lower triangularity of trigonometric
Calogero-Moser interactions which are necessary for establishing
quasi-exact solvability of trigonometric Inozemtsev models.

\section{Basic tool and Classical Inozemtsev models}
\label{inomode}
\setcounter{equation}{0}

\subsection{Basic tool}

One basic tool for showing the existence of some exact eigenfunctions
(quasi-exact solvability) is
the following simple fact.
Let $W=W(q)$ be a real smooth function of the coordinate(s), then trivial
differentiation formulas \ ($p=-i\partial/\partial q$)
\[
    p^2 e^W=-\left[\left({\partial W\over{\partial
    q}}\right)^2+{\partial^2W\over{\partial q^2}}\right] e^W,\qquad
    \sum_{j=1}^r p_{j}^2\,
    e^W=-\sum_{j=1}^r\left[\left({\partial W\over{\partial
    q_{j}}}\right)^2+{\partial^2W\over{\partial q_{j}^2}}\right]e^W,
\]
imply that $e^W$ is an eigenfunction of the Hamiltonian
$H$ with eigenvalue 0:
\begin{eqnarray}
    &&H\,e^W=0,\nonumber\\
    &&H={1\over2}p^2+{1\over2}\left[\left({\partial W\over{\partial
    q}}\right)^2+{\partial^2W\over{\partial q^2}}\right],\!\!\quad
     H={1\over2} \sum_{j=1}^r p_{j}^2+{1\over2}\sum_{j=1}^r
     \left[\left({\partial W\over{\partial
    q_{j}}}\right)^2+{\partial^2W\over{\partial q_{j}^2}}\right],
    \label{eq:Wham}
\end{eqnarray}
so long as it is square integrable
\begin{equation}
    \int e^{2W}d^r\!q<\infty.
    \label{eq:finnorm}
\end{equation}
This is the simplest example of quasi-exact solvability.
Looked differently, one might say this is
a property of `factorised' Hamiltonians
\begin{equation}
    H={1\over2}\left(p-i{\partial W\over{\partial
    q}}\right)\left(p+i{\partial W\over{\partial
    q}}\right),\qquad
      H={1\over2} \sum_{j=1}^r\left(p_{j}-i{\partial W\over{\partial
    q_{j}}}\right)\left(p_{j}+i{\partial W\over{\partial
    q_{j}}}\right),
    \label{facHam}
\end{equation}
together with a differential operator(s) that annihilate the state
\begin{equation}
\left(p+i{\partial W\over{\partial
    q}}\right)\,e^W=0,\qquad \left(p_{j}+i{\partial W\over{\partial
    q_{j}}}\right)\,e^W=0, \quad j=1,\ldots, r.
\end{equation}
This fact can also be considered as the very base of supersymmetric
quantum mechanics.

This gives the ground state
wavefunction of the quantum Calogero-Moser models.
In other words, all the (quantum integrable) Calogero-Moser models can be
described by pre-superpotentials $W$ \cite{OPq,bms}.
To sum up, if a Hamiltonian can be expressed in terms of $W$
as (\ref{eq:Wham}) or (\ref{facHam}) up to a constant, the existence of
one eigenfunction is guaranteed save the square normalisability.
Throughout this paper we call function $W$  a {\em pre-superpotential\/}.
\subsection{Classical models}
Here we present classical
Inozemtsev models together with Calogero-Moser models for comparison.
The dynamical variables are canonical coordinates $\{q_{j}\}$ and their
canonical conjugate momenta $\{p_j\}$.
We denote them by $r$-dimensional
vectors $q$ and $p$ with standard inner product:
\begin{displaymath}
 q=(q_{1},\cdots,q_{r})\in {\bf R}^r,\quad
 p=(p_1,\cdots,p_r)\in {\bf R}^r,\quad
 p^2=p\cdot p=p_{1}^2+\cdots+p_{r}^2.
\end{displaymath}
As is well-known for a root system $\Delta$ (rank $r$) and four types
of potentials, elliptic, trigonometric, hyperbolic and rational,
Calogero-Moser and Inozemtsev models are classically completely integrable.
In this
paper we discuss only those models based on  classical root  systems,
that is $A$
type and
$BC$ (and $D$) type and {\em degenerate}  potentials, that is trigonometric,
hyperbolic and rational potentials. Since algebraic structures are almost
the same for the trigonometric and  hyperbolic potential models, we discuss
trigonometric case as a  representative.
Among various types of Inozemtsev models \cite{bib:ad77}-\cite{Ino0}
we focus our attention to the {\em supersymmetrisable\/} models,
namely to those
models whose Hamiltonians  can be collectively
expressed in terms of a pre-superpotential $W=W(q)$ \cite{bms} as
\begin{equation}
 H={1\over2}p^2+{1\over2}\sum_{j=1}^r
     \left({\partial W\over{\partial
    q_{j}}}\right)^2,
\label{susyla1}
\end{equation}
or  `factorisable' at the classical level:
\begin{equation}
H={1\over2}\sum_{j=1}^r
     \left(p_{j}-i{\partial W\over{\partial
    q_{j}}}\right) \left(p_{j}+i{\partial W\over{\partial
    q_{j}}}\right).
\label{susyla2}
\end{equation}
Each specific model in this class is given by a choice of $W$, which are
listed as below.

\subsection{$BC$ type Calogero-Moser models}\label{subsec.BCrCal-Mo}
The rational model pre-superpotential  $W$
\begin{equation}
W_{\mbox{\scriptsize
C-M}}=g_{M}\sum_{j<k}^r\left\{\log|q_{j}-q_{k}|+\log|q_{j}+q_{k}|\right\}
    +g_{S}\sum_{j=1}^r\log|q_{j}|,
    \label{eq:Bcrat}
\end{equation}
contains two real coupling constants $g_{M}$ for the middle
roots (length${}^2$=2) and $g_{S}$ for the short roots
(length${}^2$=1).
The trigonometric model pre-superpotential $W$
\begin{eqnarray}
    W_{{\mbox{\scriptsize
C-M}}}&=&g_{M}\sum_{j<k}^r\left\{\log|\sin(q_{j}-q_{k})|
    +\log|\sin(q_{j}+q_{k})|\right\}\nonumber\\
    &&+g_{S}\sum_{j=1}^r\log|\sin q_{j}|+g_{L}\sum_{j=1}^r\log|\sin 2q_{j}|,
    \label{eq:Bctrig}
\end{eqnarray}
has one more  coupling constant than the rational case,
$g_{L}$ for the long roots (length${}^2$=4).
For the rational potential, the long roots and
short roots are essentially the same.
When $g_{L}=0$ and
$g_{S}=0$ the models belong to the $D_{r}$ root system.
If  $g_{L}=0$ and
$g_{S}\neq0$ ( $g_{L}\neq0$ and
$g_{S}=0$) the models belong to the $B_{r}$ ($C_{r}$) root system.
Throughout this paper we put the scale factor in the trigonometric functions
to
unity for simplicity.

\subsection{$A$ type Calogero-Moser models}\label{subsec.ArCal-Mo}
For  $A$ type models, it is customary to express the roots by
embedding in a space with one higher dimensions. We will discuss
$A_{r-1}$ models with $r$ degree of freedom. Since all the roots have
the same length, $A$ type models have only one real coupling constant
$g$.
The rational model pre-superpotential $W$ is given by
\begin{equation}
    W_{{\mbox{\scriptsize
C-M}}}=g\sum_{j<k}^r\log|q_{j}-q_{k}|,
    \label{eq:Arat}
\end{equation}
whereas the trigonometric pre-superpotential $W$ reads
\begin{equation}
    W_{{\mbox{\scriptsize
C-M}}}=g\sum_{j<k}^r\log|\sin(q_{j}-q_{k})|.
    \label{eq:Atrig}
\end{equation}

\subsection{$BC$ type Inozemtsev models }\label{subsec.BCrIno}

The rational supersymmetrisable $BC$ type
Inozemtsev model \cite{bib:le-wo84,Ino0}
has two more real coupling  constants, $a$ and $b$, than the
corresponding Calogero-Moser model,
\begin{equation}
    W=-\sum_{j=1}^r\left({a\over4}q_{j}^4
    +{b\over2}q_{j}^2\right)+W_{{\mbox{\scriptsize
C-M}}}(\ref{eq:Bcrat}),
    \label{eq:BcIn0rat}
\end{equation}
which leads to degree six polynomial potentials.
The trigonometric supersymmetrisable $BC$ type Inozemtsev model
\cite{bib:le-wo84,Ino0} has also two more real coupling
constants, $a$ and $b$,
than the corresponding Calogero-Moser model,
\begin{equation}
    W=\sum_{j=1}^r\left(-{a\over2}\cos2q_{j}
    +{b\over2}\log|\cot q_{j}|\right)+W_{{\mbox{\scriptsize
C-M}}}(\ref{eq:Bctrig}).
    \label{eq:BcIn0tri}
\end{equation}

\subsection{$A$ type Inozemtsev models}\label{subsec.ArIno}

The rational supersymmetrisable $A$ type Inozemtsev model
\cite{bib:ad77}-\cite{bib:in84} has two more real coupling
constants, $a$ and $b$, than the
corresponding Calogero-Moser model,
\begin{equation}
    W=\sum_{j=1}^r\left({a\over3}q_{j}^3
    +{b\over2}q_{j}^2\right)+W_{{\mbox{\scriptsize
C-M}}}(\ref{eq:Arat}),
    \label{eq:AIn0rat}
\end{equation}
which leads to degree four polynomial potentials.
The trigonometric $A$ type Inozemtsev model
\cite{bib:ad77}-\cite{bib:in84} has
only one more real coupling  constant, $a$ , than the
corresponding Calogero-Moser model,
\begin{equation}
    W=\sum_{j=1}^r\left(-{a\over2}\cos2q_{j}
    \right)+W_{{\mbox{\scriptsize
C-M}}}(\ref{eq:Atrig}).
    \label{eq:AIn0trig}
\end{equation}

It should be emphasised that these additional interactions of the
supersymmetrisable Inozemtsev models are all `single particle' type.
In the following three sections \S\ref{sec.oneBCrat}-\S\ref{sec.Ratone}
we will investigate the characteristic single particle dynamics of
supersymmetrisable Inozemtsev models.

\section{Rational $BC$ type Inozemtsev model with one degree of
freedom}\label{sec.oneBCrat}
\setcounter{equation}{0}

Let us start with a pre-superpotential $W=W(q)$ which is decomposed into
two parts:
\begin{eqnarray}
    W & = & W_{0}+W_{1},
    \label{eq:wsum}  \\
     &  & W_{0}=-\left({a\over4}q^4
    +{b\over2}q^2\right)+g_{S}\log|q|,\quad a>0,\quad g_{S}>0,
    \label{eq:sexpot}  \\
     &  & W_{1}=\sum_{k=1}^{\mathcal M}\log|q^2-\xi_{k}|,
    \label{eq:add}
\end{eqnarray}
in which ${\mathcal M}$ is an arbitrary non-negative integer and
$\{\xi_{k}\}$'s are distinct but as yet undetermined parameters.
The first part $W_{0}$ corresponds to the `single particle
interactions' of the $BC$ type rational Inozemtsev model, which is an even
function  of $q$. The added part gives rise to an arbitrary polynomial in
$q^2$ of  degree ${\mathcal M}$ in $e^W=e^{W_{0}}\prod_{k=1}^{\mathcal
M}(q^2-\xi_{k})$.
Since
\begin{displaymath}
\left({\partial W\over{\partial
    q}}\right)^2+{\partial^2W\over{\partial q^2}}=\left({\partial
    W_{0}\over{\partial
    q}}\right)^2+{\partial^2W_{0}\over{\partial q^2}}+2{\partial
    W_{0}\over{\partial
    q}}{\partial
    W_{1}\over{\partial
    q}}  +\left({\partial
    W_{1}\over{\partial
    q}}\right)^2+{\partial^2W_{1}\over{\partial q^2}},
\end{displaymath}
we will evaluate the terms containing $W_{1}$:
\begin{eqnarray}
     &  & 2{\partial
    W_{0}\over{\partial
    q}}{\partial
    W_{1}\over{\partial
    q}}  +\left({\partial
    W_{1}\over{\partial
    q}}\right)^2+{\partial^2W_{1}\over{\partial q^2}}
    \nonumber  \\
     &  & =2\left\{-(aq^3+bq)+{g_{S}\over{q}}\right\}
     \sum_{k=1}^{\mathcal M}{2q\over{q^2-\xi_{k}}}
    +\sum_{k=1}^{\mathcal M}{2\over{q^2-\xi_{k}}}
     +8q^2\sum_{k<l}{1\over{q^2-\xi_{k}}}{1\over{q^2-\xi_{l}}}.
    \label{eq:W1terms}
\end{eqnarray}
This is a meromorphic function in $q^2$ with at most simple poles.
We demand that the residues of the simple poles, $q^2=\xi_{k}$,
$k=1,\ldots,{\mathcal M}$ should all vanish \cite{muw},
which  results in a
set of rational (``Bethe ansatz'' type) equations for $\{\xi_{k}\}$'s:
\begin{equation}
    2\left\{-(a\xi_{k}^2+b\xi_{k})+g_{S}\right\}+1+4\xi_{k}\sum_{l\neq
    k}{1\over{\xi_{k}-\xi_{l}}}=0,\quad k=1,\ldots,{\mathcal M}.
    \label{eq:beteq}
\end{equation}
Then expression (\ref{eq:W1terms}) becomes a linear polynomial in
$q^2$, which is easy to evaluate
\begin{displaymath}
        (\ref{eq:W1terms})=-4a{\mathcal M}q^2-4bM-4a\sum_{k=1}^{\mathcal
           M}\xi_{k}.
           \label{eq:linterm}
\end{displaymath}
Thus we arrive at
\begin{eqnarray}
    \left({\partial W\over{\partial
    q}}\right)^2+{\partial^2W\over{\partial q^2}} & = &
    \left({\partial
    W_{0}\over{\partial
    q}}\right)^2+{\partial^2W_{0}\over{\partial q^2}}-4a{\mathcal
    M}q^2-2E_{1},
    \label{eq:wsq}  \\
     E_{1}& = & 2bM+2a\sum_{k=1}^{\mathcal
    M}\xi_{k}.
    \label{eq:enei}
\end{eqnarray}
It should be emphasised that except for the constant term $E_{1}$ the
expression (\ref{eq:wsq}) is independent of the  parameters
$\{\xi_{k}\}$'s introduced in $W_{1}$.

This means that, for each set of solutions $\{\xi_{k}\}$'s (with real
$\sum\xi_k$ and up to the  ordering), we have an eigenstate
\begin{equation}
    e^W=e^{W_{0}}\prod_{k=1}^{\mathcal
M}(q^2-\xi_{k})=q^{g_{S}}\,e^{-({a\over4}q^4
    +{b\over2}q^2)}\prod_{k=1}^{\mathcal
M}(q^2-\xi_{k})
    \label{eq:eigstat}
\end{equation}
with eigenvalue $E_{1}$, (\ref{eq:enei}), of the Hamiltonian
\begin{eqnarray}
    H & = & {1\over2}p^2+{1\over2}\left[\left({\partial W_{0}\over{\partial
    q}}\right)^2+{\partial^2W_{0}\over{\partial q^2}}\right]
-2a{\mathcal M}q^2,
    \label{eq:exham}  \\
     & = &
     {1\over2}p^2+{1\over2}q^2(aq^2+b)^2+{g_{S}(g_{S}-1)\over{2q^2}}
     -a({3\over2}+2{\mathcal M}+g_{S})q^2-{b\over2}(1+2g_{S}).
    \label{eq:exham2}
\end{eqnarray}
It has $q^6$, $q^4$, $q^2$ and $1/q^2$ potentials and a part of the
coefficients of the
quadratic potential is {\em quantised}. Because of the singular centrifugal
term $1/q^2$, we restrict the function space to the half line,
$(0,+\infty)$.
The restriction on the coupling constants, $a>0$ is for securing the
square integrability of $e^W$ at $q=+\infty$ and $g_{S}>0$ for
finiteness at $q=0$.

The above result implies that by adding a single
term $-2a{\mathcal  M}q^2$ to the Hamiltonian
the single particle $BC$ type rational Inozemtsev
model can be made {\em quasi-exactly solvable}, that is a finite
number of eigenstates together with their eigenvalues can be obtained
exactly by algebraic means.
The very term \(-{a\over4}q^4\) in \(W_0\) that obstructs quantum
integrability is instrumental for the introduction of the additional term
$-2a{\mathcal  M}q^2$.
The eigenfunction (\ref{eq:eigstat})
of the above {\em quasi-exactly solvable} system
belongs to a ``polynomial space''
\begin{equation}
    {\mathcal V}_{\mathcal M}
=\mbox{Span}\left[1,q^2,\ldots,q^{2k},\ldots,q^{2\mathcal
    M}\right]e^{W_{0}}.
    \label{eq:V0def}
\end{equation}
In other words, the Hamiltonian (\ref{eq:exham2}) leaves this
polynomial space invariant:
\begin{equation}
    H{\mathcal
    V}_{\mathcal M}\subseteq {\mathcal
    V}_{\mathcal M}.
    \label{eq:hinv}
\end{equation}
Therefore these ``polynomial space''  can be called the {\em exactly
solvable
sector} of the system. It is elementary to see that the ``polynomial space''
${\mathcal
    V}_{\mathcal M}$ is annihilated by an ${\mathcal N}={\mathcal M}+1$
st order  differential operator $P_{\mathcal N}$:
\begin{eqnarray}
     P_{\mathcal N}& = & \prod_{k=0}^{\mathcal N-1}\left(D+i{k\over
q}\right)=\left(D+{i({\mathcal N}-1)\over q}\right)\cdots\left(D+{i\over
q}\right)D,
    \label{eq:PNdef}  \\
     P_{\mathcal N}{\mathcal
    V}_{\mathcal M}& = & 0,\qquad\qquad \qquad\qquad \qquad\qquad
\qquad\quad
     D=p+i{\partial W_{0}\over{\partial q}}.
    \label{eq:annv}
\end{eqnarray}
Since $P_{\mathcal N}$ is an ${\mathcal N}={\mathcal M}+1$ st order
differential operator, it is obvious that ${\mathcal
    V}_{\mathcal M}$ gives the entire solution space of a differential
equation
\[
P_{\mathcal N}y=0.
\]
This differential operator together with its hermitian conjugate
defines a {\em higher derivative} \cite{ais} or {\em non-linear}
\cite{plyu} or an ${\mathcal N}$-{\em fold} \cite{aoy} supersymmetry
 generated by
\begin{equation}
    Q=P_{\mathcal N}\psi^\dagger,\quad Q^\dagger=P_{\mathcal
    N}^\dagger\psi,
    \label{eq:QNdef}
\end{equation}
in which $\psi$ and $\psi^\dagger$ are fermion annihilation and
creation operators. The ``polynomial space'' ${\mathcal
    V}_{\mathcal M}$ is characterised as the zero-modes of $Q$ and
$Q^\dagger$
\begin{equation}
     Q{\mathcal  V}_{\mathcal M}=Q^\dagger{\mathcal  V}_{\mathcal M}=0,
     \label{eq:Qanni}
\end{equation}
which is the generalisation of the property of the ground state of the
ordinary (${\mathcal N}=1$) supersymmetric quantum mechanics. The
second  equality \(Q^\dagger{\mathcal  V}_{\mathcal M}=0\) is trivial
since
\({\mathcal  V}_{\mathcal M}\) has zero fermion number.

The structure of the {\em exactly solvable sector} can be better
understood by making a similarity transformation of $H$ by
$e^{W_{0}}$ (see \cite{kps} for example):
\begin{eqnarray}
    \widetilde{H} & = & e^{-W_{0}}H\,e^{W_{0}}
    ={1\over2}p^2-{\partial W_{0}\over{\partial
     q}}{\partial\over{\partial q}}-2a{\mathcal M}q^2,
    \nonumber  \\
     & = & {1\over2}p^2+(aq^3+bq-{g_{S}\over  q})
    {\partial\over{\partial q}}
     -2a{\mathcal M}q^2.
    \label{eq:simtr}
\end{eqnarray}
Then the above ``polynomial space'' (\ref{eq:V0def}) is mapped to a
genuine polynomial space
\begin{equation}
    \widetilde{\mathcal
    V}_{\mathcal
M}=\mbox{Span}\left[1,q^2,\ldots,q^{2k},\ldots,q^{2\mathcal
    M}\right],
    \label{eq:tildeV0def}
\end{equation}
whose invariance under $\widetilde{H}$ (\ref{eq:simtr})
\[
\widetilde{H} \widetilde{\mathcal
    V}_{\mathcal M}\subseteq \widetilde{\mathcal
    V}_{\mathcal M}
\]
is rather elementary to verify. If one substitutes an expansion
\[
\widetilde{\Psi}=\sum_{k=0}^{\mathcal M}\alpha_{k}q^{2k},\quad \alpha_{0}=1
\]
into the eigenvalue equation
\[
\widetilde{H}\widetilde{\Psi}=E\widetilde{\Psi}
\]
one obtains a three term recursion
relation for
$\{\alpha_{k}\}$'s
\begin{equation}
    (k+1)(2k+1+g_{S})\alpha_{k+1}=(2kb-E)\alpha_{k}+2a(k-{\mathcal
    M}-1)\alpha_{k-1}.
    \label{eq:bendu}
\end{equation}
This determines $\alpha_{k}$ as a polynomial in $E$ of degree $k$
which is a Bender-Dunne polynomial \cite{bender} in the naivest sense.
The condition $\alpha_{\mathcal M+1}=0$
gives the characteristic equation of $\widetilde{H}$:
\begin{equation}
 \alpha_{\mathcal M+1}=0 \Leftrightarrow (2{\mathcal M}b-E)\alpha_{\mathcal
 M}-2a\alpha_{\mathcal M-1}=0 \Leftrightarrow \
 \mbox{det}(\widetilde{H}-E)=0.
    \label{eq:chareq}
\end{equation}

\bigskip
In the rest of this section let us discuss the relationship between
quasi-exact solvability and ${\mathcal N}$-fold supersymmetry in the
general context. This applies to the other cases discussed in the
following sections as well.
The {\em exactly solvable sector\/} of a quasi-exactly solvable theory is
characterised by its ``polynomial space''
\begin{equation}
    {\mathcal V}_{\mathcal
M}=\mbox{Span}\left[1,h,\ldots,h^{k},\ldots,h^{\mathcal
    M}\right]e^{W_{\mbox{\scriptsize gen}}},
    \label{eq:VGdef}
\end{equation}
which is invariant under Hamiltonian
\begin{equation}
    H_{\mbox{\scriptsize gen}}{\mathcal V}_{\mathcal M}\subseteq{\mathcal
V}_{\mathcal M}.
    \label{eq:HGinv}
\end{equation}
In these formulas the subscript ``gen" in $H$ and $W$ stands
for  `generic' and the function $h=h(q)$ need not be a polynomial in $q$.
(For example, in the trigonometric $BC$ type Inozemtsev model (see
section \ref{sec.Trigone}) $h(q)=\sin^2q$.)
It is straightforward to verify that
the ``polynomial space'' ${\mathcal  V}_{\mathcal M}$ is annihilated by
an
${\mathcal N}={\mathcal M}+1$ st order
differential operator $P_{\mathcal N}$:
\begin{eqnarray}
     P_{\mathcal N}& = & \prod_{k=0}^{\mathcal N-1}(D_{\mbox{\scriptsize
gen}}+i{kE(q)}),\quad
     D_{\mbox{\scriptsize gen}}=p+i{\partial W_{\mbox{\scriptsize
gen}}\over{\partial q}},\quad
     E(q)\equiv{h''(q)\over{h'(q)}},
    \label{eq:PNGdef}  \\
     P_{\mathcal N}{\mathcal
    V}_{\mathcal M}& = & 0.
    \label{eq:VGannv}
\end{eqnarray}
As above, the ``polynomial space'' ${\mathcal  V}_{\mathcal M}$ gives the
entire solution space of the differential equation \(P_{\mathcal
N}y=0\). One could summarise the situation as {\em the exactly solvable
sector of a quasi-exactly solvable dynamics is characterised as the
states annihilated\/} (\ref{eq:Qanni})
{\em by the generators $Q$ and $Q^\dagger$\/} (\ref{eq:QNdef}) {\em of an
${\mathcal N}$-fold supersymmetry.\/}

On the other hand, let us suppose that one has a pair of
Hamiltonians $H_{\mbox{\scriptsize gen}}$ and $H_{\mbox{\scriptsize gen}}^+$
which are intertwined by $P_{\mathcal N}$ \cite{aoy,ddt,jsuz}:
\begin{equation}
   P_{\mathcal N} H_{\mbox{\scriptsize gen}}-H_{\mbox{\scriptsize
gen}}^+P_{\mathcal N}=0.
    \label{eq:intrel}
\end{equation}
Let ${\mathcal V}_{\mathcal M}$ be the space of solutions of the
differential  equation $ P_{\mathcal N}y=0$, which is finite dimensional.
Then  from (\ref{eq:intrel}) one obtains $P_{\mathcal N}
H_{\mbox{\scriptsize
gen}}{\mathcal  V}_{\mathcal M}=0$ and thus deduces that the finite
dimensional
space
${\mathcal V}_{\mathcal M}$ is
invariant under $H_{\mbox{\scriptsize gen}}$:\
\(
H_{\mbox{\scriptsize gen}}
{\mathcal V}_{\mathcal M}\subseteq{\mathcal V}_{\mathcal M}
\), (\ref{eq:HGinv}).
One could summarise this as {\em the quasi-exact solvability of\/}
$H_{\mbox{\scriptsize gen}}$
{\em is a consequence of the ${\mathcal N}$-fold supersymmetry and the
intertwining relation \/} (\ref{eq:intrel}).
The spectral equivalence of $H_{\mbox{\scriptsize gen}}$ and
$H_{\mbox{\scriptsize gen}}^+$ holds outside of ${\mathcal V}_{\mathcal M}$
as in the ordinary (${\mathcal N}=1$) supersymmetric quantum mechanics.

\section{Trigonometric $BC$ type Inozemtsev model with one degree of
freedom}\label{sec.Trigone}
\setcounter{equation}{0}
Here we consider a one dimensional quantum mechanical system with the
following super potential $W$ in a finite interval $[0,\pi/2]$:
\begin{eqnarray}
    W & = & W_{0}+W_{1},
    \label{eq:wsum2}  \\
     &  & W_{0}=-{a\over2}\cos2q
    +{b\over2}\log|\cot q|+g_{S}\log|\sin q|,
    \label{eq:doubsg}  \\
    & & \qquad \quad  g_{S}>0,\quad g_{S}>{b\over2}>0,
    \label{gsglb}\\
     &  & W_{1}=\sum_{k=1}^{\mathcal M}\log|\sin^2q-\xi_{k}|,
    \label{eq:add2}
\end{eqnarray}
in which $W_0$ part is obtained by retaining the single particle part of the
trigonometric $BC$ type Inozemtsev model (\ref{eq:BcIn0tri}).
It is an even function of $q$ and it reduces to a well-known
``double sine-Gordon"
quantum mechanics \cite{razavy} if only the first term
${a\over2}\cos2q$ is kept.
Here we have not included the long root term $g_{L}\log|\sin 2q|$ in
(\ref{eq:Bctrig}) since it can be expressed as a linear combination of
$\log|\cot q|$ and $\log|\sin q|$ terms.
 As in the
previous section, we evaluate the terms containing
$W_1$ in
$(\partial W/\partial  q)^2+\partial^2W/\partial q^2$:
\begin{eqnarray}
&&2\left(a\sin2q-{b\over{\sin2q}}+g_S\cot q\right)
\sum_{k=1}^{\mathcal M}{2\sin q\cos q\over{\sin^2q-\xi_k}}\nonumber\\
&& +8\sin^2q\cos^2q\sum_{k<l}{1\over{\sin^2q-\xi_{k}}}
{1\over{\sin^2q-\xi_{l}}}+2\cos2q\sum_{k=1}^{\mathcal M}
{1\over{\sin^2q-\xi_{k}}},
\label{bctriw1}
\end{eqnarray}
which is a meromorphic function in $x=\sin^2q$ with at most simple  poles:
\begin{eqnarray}
(\ref{bctriw1})&=&2\left(\phantom{\mbox{\Large I}}
\hspace{-2mm}4ax(1-x)-b+2g_S(1-x)\right)
\sum_{k=1}^{\mathcal M}{1\over{x-\xi_{k}}}\nonumber\\
&& + 8x(1-x)\sum_{k<l}{1\over{x-\xi_{k}}}
{1\over{x-\xi_{l}}}+2(1-2x)\sum_{k=1}^{\mathcal M}{1\over{x-\xi_{k}}}.
\label{bcrxex}
\end{eqnarray}
As in the previous case we demand that the
residues at the simple poles $x=\xi_k$,
$k=1,\ldots,{\mathcal M}$ should all vanish. This requires that
$\{\xi_k\}$'s
should obey a set of rational equations
\begin{eqnarray}
(4a\xi_k+2g_S)(1-\xi_k)-b+1-2\xi_k+4\xi_k(1-\xi_k)\sum_{l\neq
k}{1\over{\xi_k-\xi_l}}=0,\quad
k=1,\ldots,{\mathcal M}.\label{sinbet}
\end{eqnarray}
Then expression (\ref{bcrxex}) becomes a linear function in $x=\sin^2q$
which is easy to
evaluate
\[
(\ref{bctriw1})=-8a{\mathcal M}\sin^2q
-8a\sum_{k=1}^{\mathcal M}\xi_k-4{\mathcal
M}(g_S+{\mathcal M}).
\]
Thus we arrive at
\begin{eqnarray}
    \left({\partial W\over{\partial
    q}}\right)^2+{\partial^2W\over{\partial q^2}} & = &
    \left({\partial
    W_{0}\over{\partial
    q}}\right)^2+{\partial^2W_{0}\over{\partial q^2}}-8a{\mathcal
    M}\sin^2q-2E_{1},
    \label{eq:wsq2}  \\
     E_{1}& = & 4a\sum_{k=1}^{\mathcal
    M}\xi_{k}+2{\mathcal M}(g_S+{\mathcal M}).
    \label{eq:enei2}
\end{eqnarray}
Again, except for the constant term the expression (\ref{eq:wsq2})
 is independent of the
parameters $\{\xi_k\}$'s. Thus for each real solution of (\ref{sinbet}),
we have an
eigenfunction
\begin{eqnarray}
e^W&=&e^{W_0}\prod_{k=1}^{\mathcal M}(\sin^2q-\xi_k)=(\sin q)^{g_S}(\cot
q)^{b\over2}e^{-{a\over2}\cos2q}\prod_{k=1}^{\mathcal M}(\sin^2q-\xi_k),
\label{onetrisol}
\end{eqnarray}
with eigenvalue $E_1$, (\ref{eq:enei2}), of the Hamiltonian
\begin{eqnarray}
    H & = & {1\over2}p^2+{1\over2}\left[\left({\partial W_{0}\over{\partial
    q}}\right)^2+{\partial^2W_{0}\over{\partial q^2}}\right]-4a{\mathcal
    M}\sin^2q.
    \label{eq:exham3}
\end{eqnarray}
The eigenfunction (\ref{onetrisol}) is square integrable
in \([0,\pi/2]\) due to the
restriction on the parameters (\ref{gsglb}).
In other words, the above Hamiltonian (\ref{eq:exham3})
is quasi-exactly solvable.
The eigenfunction (\ref{onetrisol})
belongs to a ``polynomial space"
\begin{equation}
{\mathcal V}_{\mathcal M}=\mbox{Span}\left[1,\sin^2 q,\ldots,(\sin
q)^{2k},\ldots,(\sin q)^{2\mathcal M}\right]e^{W_0},
\end{equation}
which is invariant under the action of the Hamiltonian
$H{\mathcal V}_{\mathcal M}\subseteq{\mathcal V}_{\mathcal M}$.
It is easy to see that  ${\mathcal
    V}_{\mathcal M}$ is annihilated by an ${\mathcal N}={\mathcal M}+1$
st order  differential operator $P_{\mathcal N}$:
\begin{eqnarray}
     P_{\mathcal N}& = & \prod_{k=0}^{\mathcal N-1}(D+i{2k\cot2q}),\quad
     D=p+i{\partial W_{0}\over{\partial q}},
    \label{eq:PNdef2}  \\
     P_{\mathcal N}{\mathcal
    V}_{\mathcal M}& = & 0.
    \label{eq:annv2}
\end{eqnarray}
Thus the general statements in the previous section concerning the
quasi-exact
solvability and ${\mathcal N}$-fold supersymmetry also hold in this case.

In order to investigate the structure of the {\em exactly solvable sector}
we make as before the similarity transformation of $H$ by
$e^{W_{0}}$:
\begin{eqnarray}
    \widetilde{H} & = & e^{-W_{0}}H\,e^{W_{0}}
      =  {1\over2}p^2-{\partial W_{0}\over{\partial
     q}}{\partial\over{\partial q}}-4a{\mathcal M}\sin^2q,
\nonumber  \\
     & = & {1\over2}p^2-(a\sin2q-{b\over{\sin2q}}+g_{S}\cot q)
{\partial\over{\partial q}}
     -4a{\mathcal M}\sin^2q.
    \label{eq:simtr2}
\end{eqnarray}
The eigenfunction of $\widetilde{H}$ in the polynomial space
$\widetilde{\mathcal V}_{\mathcal M}=\mbox{Span}\left[1,\sin^2
q,\ldots,(\sin q)^{2\mathcal M}\right]$,
can be obtained by substituting an expansion
\(
\widetilde{\Psi}=\sum_{k=0}^{\mathcal M}\alpha_{k}(\sin q)^{2k}\),
\(\alpha_{0}=1
\)
into the eigenvalue equation
\(
\widetilde{H}\widetilde{\Psi}=E\widetilde{\Psi}
\).
This again leads to a three term recursion relation for
``Bender-Dunne" polynomials
\(\{\alpha_k(E)\}\)'s:
\begin{equation}
(k+1)(2k+1-b+2g_S)\alpha_{k+1}=
\left(2k(1-2a+2g_S)-E\right)\alpha_k
+4a(k-{\mathcal
M}-1)\alpha_{k-1}.
\end{equation}
Again $\alpha_k(E)$ is a polynomial in $E$ of degree $k$ and the condition
$\alpha_{\mathcal M+1}=0$ gives the characteristic equation for
$\widetilde{H}$:
\[
\alpha_{\mathcal M+1}=0 \Leftrightarrow \left(2{\mathcal
M}(1-2a+2g_S)-E\right)\alpha_{\mathcal
 M}-4a\alpha_{\mathcal M-1}=0 \Leftrightarrow \
 \mbox{det}(\widetilde{H}-E)=0.
\]

\section{Rational $A$ type Inozemtsev model with one degree of
freedom}\label{sec.Ratone}
\setcounter{equation}{0}
This is an interesting example which fails to achieve quasi-exact
solvability
due to the lack of square integrability of the eigenfunction.
It is interesting to know how far the algebraic procedures go
in parallel with the
previous cases. We start with the following pre-superpotential $W$ which is
obtained  by retaining the single particle part of $W$ in
(\ref{eq:AIn0rat}):
\begin{eqnarray}
    W  =  W_{0}+W_{1},
    \quad W_{0}={a\over3}q^3
    +{b\over2}q^2,
    \quad W_{1}=\sum_{k=1}^{\mathcal M}\log|q-\xi_{k}|.
    \label{eq:add3}
\end{eqnarray}
This is cubic in $q$ and leads to a quartic potential of $q$, see
(\ref{eq:exhamrat}).  The terms containing $W_1$ in
$(\partial W/\partial  q)^2+\partial^2W/\partial q^2$ are:
\begin{equation}
2(aq^2+bq)\sum_{k=1}^{\mathcal
M}{1\over{q-\xi_k}}+2\sum_{k<l}{1\over{q-\xi_k}}{1\over{q-\xi_l}}.
\label{w1A}
\end{equation}
From the requirement of vanishing residue at $q=\xi_k$, we obtain
rational equations
\begin{equation}
a\xi_k^2+b\xi_k+\sum_{l\neq k}{1\over{\xi_k-\xi_l}}=0,
\quad k,l=1,\ldots,{\mathcal M},
\label{ratAeq}
\end{equation}
and the expression (\ref{w1A}) reads
\[
(\ref{w1A})=2a{\mathcal M}q+2a\sum_{k=1}^{\mathcal M}\xi_k+2b{\mathcal M}.
\]
Thus for each real solution $\{\xi_k\}$ of (\ref{ratAeq})
we obtain an ``eigenfunction"
\begin{equation}
    e^W=e^{W_{0}}\prod_{k=1}^{\mathcal
M}(q-\xi_{k})=e^{({a\over3}q^3
    +{b\over2}q^2)}\,\prod_{k=1}^{\mathcal
M}(q-\xi_{k})
    \label{eq:eigstatratA}
\end{equation}
of a Hamiltonian
\begin{eqnarray}
    H & = & {1\over2}p^2+{1\over2}\left[\left({\partial W_{0}\over{\partial
    q}}\right)^2+{\partial^2W_{0}\over{\partial q^2}}\right]
    +a{\mathcal M}q,
    \nonumber  \\
     & = &
     {1\over2}p^2+{1\over2}q^2(aq+b)^2+
     a({\mathcal M}+{1\over2})q + {b\over2},
    \label{eq:exhamrat}
\end{eqnarray}
with energy
\begin{equation}
E=-a\sum_{k=1}^{\mathcal M}\xi_k-b{\mathcal M}.
\label{harmener}
\end{equation}
Let us recall the simple facts about the limiting case of
$a=0$ and $b=-\omega$, $\omega>0$.
Then the Hamiltonian (\ref{eq:exhamrat}) becomes
that of the simple harmonic oscillator
with angular frequency $\omega$ and the equations (\ref{ratAeq}) determine
$\{\xi_k\}$'s as the zeros of Hermite polynomials \cite{stie}, with scaling
by
$\sqrt{\omega}$. This results in the well-known eigenfunction with
Hermite polynomials
(\ref{eq:eigstatratA}) and the spectrum $E=\omega{\mathcal M}$
(\ref{harmener}). (The
zero-point energy
$\omega/2$ is contained in the Hamiltonian (\ref{eq:exhamrat}).)
Thus at least for $b<0$ and $|a/b|\ll1$, that is the quartic and
the accompanying cubic
terms in the potential can be considered as  `perturbations',
it is expected that
the equations (\ref{ratAeq}) have real solutions and the above solution
generating method would work. However, the ``eigenfunction"
(\ref{eq:eigstatratA}) is not square integrable in the region
$(-\infty,+\infty)$ for whichever choice of the sign of $a\neq0$.
One might be
tempted to restrict the region to a half line, say  $(0,+\infty)$, by
introducing a singular potential at the origin, for example,  by adding a
term
$g\log|q|$ to $W_0$, as in the example in section
\ref{sec.oneBCrat}. However, this cannot remedy the situation because of the
wrong parity of the $aq^3$ term.

The ${\mathcal N}={\mathcal M}+1$ st order
differential operator $P_{\mathcal N}$ annihilating
the ``polynomial space''
\begin{equation}
    {\mathcal
    V}_{\mathcal M}=\mbox{Span}\left[1,q,\ldots,q^{k},\ldots,q^{\mathcal
    M}\right]e^{W_{0}},
    \label{eq:V0defratA}
\end{equation}
has a very simple form:
\begin{equation}
     P_{\mathcal N} = D^{\mathcal N},\quad
     D=p+i{\partial W_{0}\over{\partial q}}.
    \label{eq:PNdefratA}
\end{equation}
In this case the ${\mathcal N}$-fold supersymmetry generated by
$Q= P_{\mathcal N}\psi^\dagger$ and $Q^\dagger=
P_{\mathcal N}^\dagger\psi$ is {\em
spontaneously broken\/} for $a\neq0$.

The difference in the algebraic structure
from the quasi-exactly solvable case discussed
in section \ref{sec.oneBCrat} becomes clearer
by making the similarity transformation of $H$ by
$e^{W_{0}}$ :
\begin{eqnarray}
    \widetilde{H} & = & e^{-W_{0}}H\,e^{W_{0}}
    ={1\over2}p^2-{\partial W_{0}\over{\partial
     q}}{\partial\over{\partial q}}+a{\mathcal M}q,
    \nonumber  \\
     & = & {1\over2}p^2-(aq^2+bq) {\partial\over{\partial q}}
     +a{\mathcal M}q.
    \label{eq:simtrratA}
\end{eqnarray}
This maps $q^k$ to $q^{k+1}$, $q^k$ and $q^{k-2}$:
\[
\widetilde{H}q^k=a({\mathcal M}-k)q^{k+1}-bkq^k-{k(k-1)\over2}q^{k-2}.
\]
Since the last term is $q^{k-2}$ instead of
$q^{k-1}$, the three term recursion relations
for the coefficients $\{\alpha_k\}$'s in a series solution
\(
\widetilde{\Psi}=\sum_{k=0}^{\mathcal M}\alpha_{k}q^{k}\), \(\alpha_{0}=1\)
for  the eigenvalue equation
\(
\widetilde{H}\widetilde{\Psi}=E\widetilde{\Psi}
\)
do not hold any more.
The $\{\alpha_k(E)\}$'s are no longer degree $k$ polynomial in $E$.

\bigskip
We will not discuss ``trigonometric $A$ type Inozemtsev model with
one degree of  freedom", since the single particle part of
(\ref{eq:AIn0trig}) is simply
$W=-{a\over2}\cos2q$, that is the ``double sine-Gordon" quantum mechanics.
This is a well-known example of quasi-exactly solvable dynamics
\cite{razavy}
and is a special case
of the model treated in section \ref{sec.Trigone}.

\section{Rational $BC$ type Inozemtsev model }\label{sec.rBCrat}
\setcounter{equation}{0}

Following the results of the single particle case
in section \ref{sec.oneBCrat}, we
consider the following Hamiltonian
\begin{equation}
H={1\over2} \sum_{j=1}^r p_{j}^2+{1\over2}\sum_{j=1}^r
     \left[\left({\partial W_0\over{\partial
    q_{j}}}\right)^2+{\partial^2W_0\over{\partial q_{j}^2}}\right]-
2a{\mathcal  M}\sum_{j=1}^r q_j^2,
\label{mulratHam}
\end{equation}
in which ${\mathcal M}$ is an arbitrary non-negative integer and
$W_0$ is given by (\ref{eq:BcIn0rat}):
\begin{eqnarray}
    W_0&=&-\sum_{j=1}^r\left({a\over4}q_{j}^4
    +{b\over2}q_{j}^2\right)+W_{\mbox{\scriptsize C-M}},
    \label{eq:BcIn0ratagain}\\
     W_{\mbox{\scriptsize

C-M}}&=&g_{M}\sum_{j<k}^r\left\{\log|q_{j}-q_{k}|+\log|q_{j}+q_{k}|\right\}
    +g_{S}\sum_{j=1}^r\log|q_{j}|,\nonumber\\
&& a>0,\quad g_S>0,\quad g_M>0.
\label{BCrpara}
\end{eqnarray}
The Hamiltonian as well as $W_0$ are Coxeter (Weyl)
invariant of the $BC_r$ root system.
The only difference with the classical supersymmetrisable Inozemtsev model
is
the added quadratic terms proportional to ${\mathcal M}$. We will consider
the model in the principal Weyl chamber
\begin{equation}
q_1>q_2>\cdots>q_r>0.
\label{BCweyl}
\end{equation}
A special case of this Hamiltonian with one free parameter other
than ${\mathcal M}$ (plus an invisible overall scale factor) was
discussed in \cite{hs1}.

In order to show the quasi-exact solvability of
the Hamiltonian (\ref{mulratHam}) we
have to demonstrate that a certain
``exactly solvable sector"
is invariant under $H$. As before let us first define a ``polynomial space"
\begin{equation}
{\mathcal V}_{\mathcal M}
=\mbox{Span}_{0\leq n_j\leq{\mathcal M},\ 1\leq j\leq
r}\left[(q_1^2)^{n_1}\cdots(q_j^2)^{n_j}\cdots(q_r^2)^{n_r}\right]\,e^{W_0}.
\end{equation}
The ``exactly solvable sector" is the {\em permutation\/}
($q_j\leftrightarrow q_k$)
{\em invariant subspace\/} of ${\mathcal V}_{\mathcal M}$:
\begin{equation}
{\mathcal V}_{\mathcal M}^{G_{\scriptsize BC}}=
\left\{v\in{\mathcal V}_{\mathcal M}| gv=v,\ \
\forall g\in
G_{\scriptsize BC}\right\},
\end{equation}
in which $G_{\scriptsize BC}$ is the Coxeter (Weyl) group of
the $BC$ root system.
This fact can be seen easily, as in the single particle case, by similarity
transformation
\begin{eqnarray}
    \widetilde{H} & = & e^{-W_{0}}H\,e^{W_{0}}
    ={1\over2}\sum_{j=1}^rp_j^2-\sum_{j=1}^r{\partial W_{0}\over{\partial
     q_j}}{\partial\over{\partial q_j}}-2a{\mathcal
M}\sum_{j=1}^rq_j^2,
    \nonumber  \\
     & = & {1\over2}\sum_{j=1}^rp_j^2-\sum_{j=1}^r
{\partial W_{\mbox{\scriptsize C-M}}\over{\partial
     q_j}}{\partial\over{\partial q_j}}+\sum_{j=1}^r(aq_j^3+bq_j)
{\partial\over{\partial q_j}}
     -2a{\mathcal M}\sum_{j=1}^rq_j^2,
    \label{eq:simtrAr}\\[6pt]
\widetilde{\mathcal V}_{\mathcal M}&=&\
\mbox{Span}_{0\leq n_j\leq{\mathcal M},\ 1\leq j\leq
r}\left[(q_1^2)^{n_1}\cdots(q_j^2)^{n_j}\cdots(q_r^2)^{n_r}\right].
\end{eqnarray}
The proof of the invariance of
\(\widetilde{\mathcal V}_{\mathcal M}^{G_{\scriptsize BC}} \)
under  the
added terms
\(\sum_{j=1}^r(aq_j^3+bq_j) {\partial\over{\partial q_j}} -2a{\mathcal
M}\sum_{j=1}^rq_j^2\) is essentially the same as in the single particle
case.
As for the
Calogero-Moser part, $W_{\mbox{\scriptsize C-M}}$, it always decreases the
power
\(\sum_{j=1}^r n_j\) by one unit. The Coxeter (Weyl) invariance is necessary
and sufficient so that
the result remains a polynomial, without developing unwanted poles.
Thus the invariance
of  \(\widetilde{\mathcal V}_{\mathcal M}^{G_{\scriptsize BC}} \)
under \(\widetilde{H}\) and
the quasi-exact integrability of $H$ is proved.
It is obvious that the elements of
\({\mathcal V}_{\mathcal M}^{G_{\scriptsize BC}}\) are
square integrable since
\begin{equation}
e^{W_0}=e^{-\sum_{j=1}^r({a\over4}q_j^4+{b\over2}q_j^2)}
\prod_{j=1}^r(q_j)^{g_S}\prod_{j<k}(q_j^2-q_k^2)^{g_M}
\end{equation}
and the restriction on the parameters (\ref{BCrpara})
and the integration region
(\ref{BCweyl}) secure finiteness at infinity and at the boundaries
of the Weyl chambers.

Thus we have shown that the rational $BC$ type Inozemtsev model can be made
quasi-exactly solvable by adding properly quantised quadratic terms
\(-2a{\mathcal M}\sum_{j=1}^rq_j^2\).
The above ``polynomial space" \({\mathcal V}_{\mathcal M} \) is
annihilated by the following
$r$ different
 ${\mathcal N}={\mathcal M}+1$ st order commuting
differential operators $P_{\mathcal N}^{(j)}$:
\begin{eqnarray}
     P_{\mathcal N}^{(j)}& = & \prod_{k=0}^{\mathcal
N-1}(D_j+i{k\over{q_j}}),
    \quad
     D_j=p_j+i{\partial W_{0}\over{\partial{q_j}}},
    \label{eq:PNdefAr}  \\
     P_{\mathcal N}^{(j)}{\mathcal
    V}_{\mathcal M}& = & 0,\quad [P_{\mathcal N}^{(j)},P_{\mathcal
N}^{(k)}]=0,\quad j,k=1,\ldots,r.
    \label{eq:anrnv}
\end{eqnarray}
The \({\mathcal N}\)-fold supersymmetry is generated by
\begin{equation}
    Q=\sum_{j=1}^rP_{\mathcal N}^{(j)}\psi_j^\dagger,\quad
Q^\dagger=\sum_{j=1}^r(P_{\mathcal  N}^{(j)})^\dagger\psi_j,
    \label{eq:QNdefBCr}
\end{equation}
in which $\psi_j$ and $\psi_j^\dagger$ are the annihilation and
creation operators of the
$j$-th fermion.

\section{Trigonometric $BC$ type Inozemtsev model }
\label{sec.rBCtri}
\setcounter{equation}{0}

Following the results of the single particle case
in section \ref{sec.Trigone}, we
consider the following Hamiltonian
\begin{equation}
H={1\over2} \sum_{j=1}^r p_{j}^2+{1\over2}\sum_{j=1}^r
     \left[\left({\partial W_0\over{\partial
    q_{j}}}\right)^2+{\partial^2W_0\over{\partial q_{j}^2}}\right]-
4a{\mathcal  M}\sum_{j=1}^r \sin^2q_j,
\label{multriHam}
\end{equation}
in which ${\mathcal M}$ is an arbitrary non-negative integer and
$W_0$ is given by (\ref{eq:BcIn0tri}):
\begin{eqnarray}
    W_0&=&\sum_{j=1}^r\left(-{a\over2}\cos2q_{j}
    +{b\over2}\log|\cot q_{j}|\right)+W_{\mbox{\scriptsize C-M}},
    \label{eq:BcIn0triagain}\\
 W_{\mbox{\scriptsize
C-M}}&=&g_{M}\sum_{j<k}^r\left\{\log|\sin(q_{j}-q_{k})|
    +\log|\sin(q_{j}+q_{k})|\right\}\nonumber\\
    &&+g_{S}\sum_{j=1}^r\log|\sin q_{j}|.
\label{BCtriW0}
\end{eqnarray}
All the parameters $a$, $b$,  $g_M$ and $g_S$ are real and they
satisfy
\begin{equation}
 g_{S}>0,\quad g_{M}>0 \quad g_{S}>{b\over2}>0.
    \label{gsglbr}
\end{equation}

Let us recall here the argument in section \ref{sec.Trigone} for dropping
$g_{L}\log|\sin 2q_{j}|$ term in favour of $b\log|\cot q_{j}|$ term.
The Hamiltonian as well as $W_0$ are Coxeter (Weyl) invariant of the $BC_r$
root system.
The only difference with the classical supersymmetrisable Inozemtsev
model is the added \(\sin^2q_j\) terms proportional to ${\mathcal
M}$. We will consider the quantum mechanical model in the principal
Weyl alcove
\begin{equation}
\pi/2>q_1>q_2>\cdots>q_r>0,
\label{BCweylal}
\end{equation}
due to the periodicity and Coxeter (Weyl)  invariance of the model.
As in the single particle case, we will
demonstrate that a certain
``exactly solvable sector"
is invariant under $H$. As before let us first define a ``polynomial space"
\begin{equation}
{\mathcal V}_{\mathcal M}
=\mbox{Span}_{0\leq n_j\leq{\mathcal M},\ 1\leq j\leq
r}\left[(\sin^2q_1)^{n_1}\cdots(\sin^2q_j)^{n_j}
\cdots(\sin^2q_r)^{n_r}\right]
\,e^{W_0},
\label{BCrtripolysp}
\end{equation}
or equivalently:
\begin{equation}
{\mathcal V}_{\mathcal M}
=\mbox{Span}_{-{\mathcal M}\leq
n_j^\prime\leq{\mathcal M},\ 1\leq j\leq
r}\left[ \cos2(\sum_{j=1}^rn_j^\prime q_j)\right]
\,e^{W_0}.
\end{equation}
The ``exactly solvable sector" is the {\em permutation\/}
($q_j\leftrightarrow q_k$)
{\em invariant subspace\/} of ${\mathcal V}_{\mathcal M}$:
\begin{equation}
{\mathcal V}_{\mathcal M}^{G_{\scriptsize BC}}
=\left\{v\in{\mathcal V}_{\mathcal M}| gv=v,\ \
\forall g\in
G_{\scriptsize BC}\right\},
\end{equation}
in which $G_{\scriptsize BC}$ is the Coxeter (Weyl) group of
the $BC$ root system.
All these functions are square integrable, since the
integration region is finite and
the possible singularities in $e^{W_0}$ at the boundary (\ref{BCweylal})
\begin{equation}
e^{W_0}=e^{-{a\over2}\sum_{j=1}^r\cos2q_j}
\prod_{j=1}^r(\sin
q_j)^{g_S}(\cot q_j)^{b\over2}\prod_{k<l}^r[\sin(q_k-q_l)\sin
(q_k+q_l)]^{g_M},
\end{equation}
are taken care of by the restriction on the parameters (\ref{gsglbr}).
The quasi-exact solvability can be shown  by
the similarity transformation
\begin{eqnarray}
    \widetilde{H} & = & e^{-W_{0}}H\,e^{W_{0}}
    ={1\over2}\sum_{j=1}^rp_j^2-\sum_{j=1}^r{\partial W_{0}\over{\partial
     q_j}}{\partial\over{\partial q_j}}-4a{\mathcal M}\sum_{j=1}^r\sin^2
q_j,
    \nonumber  \\
     & = & {1\over2}\sum_{j=1}^rp_j^2-\sum_{j=1}^r
{\partial W_{\mbox{\scriptsize C-M}}\over{\partial
     q_j}}{\partial\over{\partial q_j}}-\sum_{j=1}^r(a\sin2q_j-
\!{b\over{\sin2q_j}})
{\partial\over{\partial q_j}}
     \!-4a{\mathcal M}\sum_{j=1}^r\sin^2\!q_j,
    \label{eq:simtrBCtri}\\[6pt]
\widetilde{\mathcal V}_{\mathcal M}& =&
\mbox{Span}_{0\leq n_j\leq{\mathcal M},\ 1\leq j\leq
r}\left[ 
(\sin^2q_1)^{n_1}\cdots(\sin^2q_j)^{n_j}\cdots(\sin^2q_r)^{n_r}\right],
\label{simtrigpoly}\\
&=&\mbox{Span}_{-{\mathcal M}\leq
n_j^\prime\leq{\mathcal M},\ 1\leq j\leq
r}\left[ \cos2(\sum_{j=1}^rn_j^\prime q_j)\right].
\label{cosbasis2}
\end{eqnarray}
The corresponding ``exactly solvable sector" is the {\em permutation\/}
($q_j\leftrightarrow q_k$) {\em invariant subspace\/} 
of $\widetilde{\mathcal
V}_{\mathcal M}$:
\begin{eqnarray}
\widetilde{\mathcal V}_{\mathcal M}^{G_{\scriptsize BC}}&=&
\left\{v\in\widetilde{\mathcal V}_{\mathcal M}|
gv=v,\
\
\forall g\in G_{\scriptsize BC}\right\}.
\label{cosbasis}
\end{eqnarray}
As for the Calogero-Moser part, the above Hamiltonian is  lower
triangular \cite{kps}  in the basis (\ref{phidef2}) of $\widetilde{\mathcal
V}_{\mathcal M}^{G_{\scriptsize BC}}$.
The lower triangularity is stronger than the invariance of the polynomial space
under the Hamiltonian.
An outline of the proof is given in Appendix B.
As for the added part
\[
-\sum_{j=1}^r\left(a\sin2q_j-\!{b\over{\sin2q_j}}\right)
{\partial\over{\partial q_j}}
   -4a{\mathcal M}\sum_{j=1}^r\sin^2q_j,
\]
the proof of the invariance of the polynomial space (\ref{simtrigpoly}) is 
essentially the same as in the single particle case.
Thus the quasi-exact solvability of the 
trigonometric $BC$ type Inozemtsev model is established.

The above ``polynomial space" \({\mathcal V}_{\mathcal M} \)
(\ref{BCrtripolysp}) is
 annihilated by the following
$r$ different
 ${\mathcal N}={\mathcal M}+1$ st order commuting
differential operators $P_{\mathcal N}^{(j)}$:
\begin{eqnarray}
     P_{\mathcal N}^{(j)}& = & \prod_{k=0}^{\mathcal
N-1}(D_j+2i{k\cot2q_j}),\quad
     D_j=p_j+i{\partial W_{0}\over{\partial{q_j}}},
    \label{eq:PNdefBCtri}  \\
     P_{\mathcal N}^{(j)}{\mathcal
    V}_{\mathcal M}& = & 0,\quad [P_{\mathcal N}^{(j)},P_{\mathcal
N}^{(k)}]=0,\quad j,k=1,\ldots,r.
    \label{eq:bctrinv}
\end{eqnarray}
The \({\mathcal N}\)-fold supersymmetry is generated by
\[
    Q=\sum_{j=1}^rP_{\mathcal N}^{(j)}\psi_j^\dagger,\quad
Q^\dagger=\sum_{j=1}^r(P_{\mathcal  N}^{(j)})^\dagger\psi_j,
\]
in which $\psi_j$ and $\psi_j^\dagger$ are
the annihilation and creation operators of the
$j$-th fermion.

\section{Trigonometric $A$ type Inozemtsev model }
\label{sec.rAtri}
\setcounter{equation}{0}
This has a much simpler Hamiltonian than the previous one:
\begin{equation}
H={1\over2} \sum_{j=1}^r p_{j}^2+{1\over2}\sum_{j=1}^r
     \left[\left({\partial W_0\over{\partial
    q_{j}}}\right)^2+{\partial^2W_0\over{\partial q_{j}^2}}\right]-
4a{\mathcal  M}\sum_{j=1}^r \sin^2q_j,
\label{mulAtriHam}
\end{equation}
in which ${\mathcal M}$ is an arbitrary non-negative integer and
$W_0$ is given by (\ref{eq:AIn0trig}):
\begin{eqnarray}
    W_0&=&-{a\over2}\sum_{j=1}^r\cos2q_{j} +W_{\mbox{\scriptsize C-M}},
  \qquad
 W_{\mbox{\scriptsize C-M}}=g\sum_{j<k}^r\log|\sin(q_{j}-q_{k})|.
 \label{eq:AIn0triagain}
\end{eqnarray}
There are only two real coupling constants $a$ and $g$ and we require
\begin{equation}
g>0
\end{equation}
for square integrability of the eigenfunctions of the form
(\ref{Artrigpoly}).
The Hamiltonian and $W_0$ are invariant under any transpositions
$(q_j,p_j)\leftrightarrow(q_k,p_k)$, which form the Coxeter group of
$A_{r-1}$.
Thus we consider the quantum mechanics in the fundamental Weyl alcove:
\begin{equation}
\pi>q_1>q_2>\cdots>q_r>0.
\label{Aweylal}
\end{equation}
Reflecting the simple form of the Hamiltonian, the ``exactly solvable
sector"
which is left invariant under $H$ has a simpler structure than those in the
previous cases.
Let us first define a space of truncated Fourier
series with two units:
\begin{equation}
{\mathcal V}_{\mathcal M}=\mbox{Span}_{-{\mathcal M}\leq
n_j\leq{\mathcal M},\ 1\leq j\leq
r}\left[  e^{2i\sum_{j=1}^rn_jq_j}\right]e^{W_0}.
\label{Artrigpoly}
\end{equation}
The ``exactly solvable sector" is the {\em permutation\/}
($q_j\leftrightarrow q_k$)
{\em invariant subspace\/} of ${\mathcal V}_{\mathcal M}$:
\begin{equation}
{\mathcal V}_{\mathcal M}^{G_{\scriptsize A}}
=\left\{v\in{\mathcal V}_{\mathcal M}| gv=v,\ \
\forall g\in
G_{\scriptsize A}\right\},
\label{fouspace}
\end{equation}
in which $G_{\scriptsize A}$ is the Coxeter (Weyl) group of the
$A$ type root system.
The quasi-exact solvability of the Hamiltonian (\ref{mulAtriHam}) is again
easily verified by the similarity transformation:
\begin{eqnarray}
    \widetilde{H} & = & e^{-W_{0}}H\,e^{W_{0}}
    ={1\over2}\sum_{j=1}^rp_j^2-\sum_{j=1}^r{\partial W_{0}\over{\partial
     q_j}}{\partial\over{\partial q_j}}-4a{\mathcal M}\sum_{j=1}^r\sin^2
q_j,
    \nonumber  \\
     & = & {1\over2}\sum_{j=1}^rp_j^2-\sum_{j=1}^r
{\partial W_{\mbox{\scriptsize C-M}}\over{\partial
    q_j}}{\partial\over{\partial
q_j}}-a\sum_{j=1}^r\sin2q_j {\partial\over{\partial q_j}}
     -4a{\mathcal M}\sum_{j=1}^r\sin^2\!q_j,
    \label{eq:simtrAtri}\\[4pt]
\widetilde{\mathcal V}_{\mathcal M}&=&\mbox{Span}_{-{\mathcal M}\leq
n_j\leq{\mathcal M},\ 1\leq j\leq
r}\left[  
e^{ 2i\sum_{j=1}^rn_jq_j}\right].
\label{simtrigpolyA2}
\end{eqnarray}
The lower triangularity of the Calogero-Moser part of the Hamiltonian in the
basis (\ref{phidef}) of (\ref{simtrigpolyA2}) 
was proven originally by Sutherland
\cite{Sut}. For the additional part
\[
-a\sum_{j=1}^r\sin2q_j {\partial\over{\partial q_j}}
     -4a{\mathcal M}\sum_{j=1}^r\sin^2\!q_j,
\]
the proo
f that it leaves the space of the truncated Fourier series
(\ref{simtrigpolyA2}) invariant is
rather elementary.

The above ``polynomial space" \({\mathcal V}_{\mathcal M} \)
(\ref{Artrigpoly}) is annihilated by the following
$r$ different
 ${\mathcal N}=2{\mathcal M}+1$ st order commuting
differential operators $P_{\mathcal N}^{(j)}$:
\begin{eqnarray}
     P_{\mathcal N}^{(j)}& = & \prod_{k=-\mathcal M}^{\mathcal
M}(D_j+2ik),\quad
     D_j=p_j+i{\partial W_{0}\over{\partial{q_j}}},
    \label{eq:PNdefAtri}  \\
     P_{\mathcal N}^{(j)}{\mathcal
    V}_{\mathcal M}& = & 0,\quad [P_{\mathcal N}^{(j)},P_{\mathcal
N}^{(k)}]=0,\quad j,k=1,\ldots,r.
    \label{eq:Atrinv}
\end{eqnarray}
The \({\mathcal N}\)-fold supersymmetry is generated by
\[
    Q=\sum_{j=1}^rP_{\mathcal N}^{(j)}\psi_j^\dagger,\quad
Q^\dagger=\sum_{j=1}^r(P_{\mathcal  N}^{(j)})^\dagger\psi_j,
\]
in which $\psi_j$ and $\psi_j^\dagger$ are the annihilation and
creation operators of the
$j$-th fermion.

We will not discuss the quasi-exact solvability of multiparticle rational
$A$
type Inozemtsev model, since its wavefunctions are not square
integrable as seen
in section \ref{sec.Ratone}. One could say that the ${\mathcal N}$-fold
supersymmetry is spontaneously broken in this case.

Let us summarise that the quasi-exact solvability of quantum Inozemtsev
models
discussed in sections \ref{sec.rBCrat}, \ref{sec.rBCtri}
and \ref{sec.rAtri} is a
consequence of the exact solvability of the quantum Calogero-Moser models
and
the quasi-exact solvability of the added single particle like interactions.

\section{Comments and discussion}
\label{sec.comm}
\setcounter{equation}{0}
It should be stressed that the present method for showing quasi-exact
solvability of single particle systems developed in section
\ref{sec.oneBCrat} to
\ref{sec.Ratone} does not depend on any existing methods
or criteria for QES, see
for example, \cite{turb}. As shown in section \ref{sec.Ratone} it also gives
the known exact solutions when the QES system tends to the harmonic
oscillator.

An interesting question along the line of arguments in this paper is  the
 `hierarchy problem' of quasi-exactly solvable systems, as in the completely
integrable systems. For example, the Inozemtsev models have higher conserved
quantities Tr$(L^k)$ obtained from the Lax pairs in Appendix A.
They define new classical and quantum Hamiltonian systems.
Can the quantum version of the higher members of the hierarchy
 be deformed to be quasi-exactly
solvable?

in
So far, elliptic Calogero-Moser models defied various attempts to construct
quantum theory based on a Hilbert space,
although existence of mutually commuting
operators are known for the $A$ type models \cite{Oshima}.
In analogy with the present arguments, it is quite natural to expect
 the quantum elliptic  Calogero-Moser models to be quasi-exactly solvable
\cite{FinkelAN,khare} rather
than  exactly integrable.

\section*{Acknowledgements}
\setcounter{equation}{0}
        R. S. is partially supported  by the Grant-in-aid from the
Ministry of Education, Culture, Sports, Science and Technology,
priority area (\#707) ``Supersymmetry and unified theory of elementary
particles". K. T. is partially supported  by the Grant-in-aid from the
Ministry of Education, Culture, Sports, Science and Technology,
(\#12640169).

\section*{Appendix A: Lax pairs for classical Inozemtsev models}
\setcounter{equation}{0}
\renewcommand{\theequation}{A.\arabic{equation}}
Here we present the Lax pairs for classical Inozemtsev models in the same
notation as is used in the main text as evidence for their classical
integrability. As mentioned in Introduction only certain subset of classical
Inozemtsev models can be made quasi-exactly solvable. We focus on the
supersymmetrisable Inozemtsev models
for simplicity of presentation.
For the full content of Lax  pairs of classical inozemtsev models
we refer to the original paper by Inozemtsev and Meshchryakov \cite{Ino0}
and for the universal Lax pairs for Calogero-Moser models in general, see
\cite{bcs2,bms}.

The Lax pair consists of a pair of  $2r \times 2r$ ($r$ rank) matrices
$L$ and $M$ such that the canonical equations of motion can be expressed
in a matrix form
\[
    \dot{L} = [L,M]
\]
and a sufficient  number of classical conserved quantities can be obtained
as
as the trace of powers of $L$, Tr$(L^k)$. The Hamiltonian is $H\propto$
Tr$(L^2)$.

\renewcommand{\thesubsection}{A.\arabic{subsection}}
\subsection{$BC$ type models}
The following Lax pair applies for the models presented in
subsection \ref{subsec.BCrIno} for the rational
as well as trigonometric  models
for proper choice of functions, $x$, $\nu$, etc. as listed below. The pair
of
matrices decomposes into diagonal and off-diagonal matrices:
\begin{eqnarray}
    L = P + X, \quad M = D + Y.
\end{eqnarray}
The diagonal matrices $P$ and $D$ are  of the form
\begin{eqnarray}
    P = \sum_{j=1}^r p_j(E_{j,j} - E_{j+r,j+r}), \quad
    D = \sum_{j=1}^r D_{j}(E_{j,j} + E_{j+r,j+r}),
\end{eqnarray}
in which $E_{j,k}$ is the usual matrix unit
$(E_{j,k})_{lm}=\delta_{lj}\delta_{mk}$.
The diagonal-free matrices $X$ and $Y$ have the form
\begin{eqnarray}
  X &=& ig_M\sum_{j\not= k} x(q_j - q_k)E_{j,k}
      + ig_M\sum_{j\not= k} x(q_j + q_k)E_{j,k+r}
    \nonumber\\
    && \mbox{}
      + ig_M\sum_{j\not= k} x(-q_j - q_k)E_{j+r,k}
      + ig_M\sum_{j\not= k} x(-q_j + q_k)E_{j+r,k+r}
    \nonumber \\
    && \mbox{}
      +2i\sum_j \nu(q_j)E_{j,j+r}
     -2i\sum_j \nu(q_j)E_{j+r,j},
    \\
  Y &=& ig_M\sum_{j\not= k} y(q_j - q_k)E_{j,k}
      + ig_M\sum_{j\not= k} y(q_j + q_k)E_{j,k+r}
    \nonumber \\
    && \mbox{}
      + ig_M\sum_{j\not= k} y(-q_j - q_k)E_{j+r,k}
      + ig_M\sum_{j\not= k} y(-q_j + q_k)E_{j+r,k+r}
    \nonumber \\
    && \mbox{}
      + i\sum_j \nu'(q_j)E_{j,j+r}
      + i\sum_j \nu'(q_j)E_{j+r,j}.
\end{eqnarray}
The diagonal elements of $D$ are given by
\begin{eqnarray}
  D_j = - ig_M\sum_{k\not= j}^r\left(z(q_j - q_k) + z(q_j + q_k)\right)
        - i\sum_{j=1}^r \tau(q_j).
\end{eqnarray}

%
Some functions are related to each other:
\begin{eqnarray}
    y(u)=dx(u)/du,\quad z(u) = x(u)^2 + {\rm const.}, \quad
    \tau(u) = 2x(2u)\nu(u) + {\rm const.}
\end{eqnarray}

The rational and trigonometric models correspond to the
following choice of functions:
\begin{enumerate}
\item
Rational model,
\begin{eqnarray}
    x(u) = \frac{1}{u}, \quad
    z(u) = \frac{1}{u^2},
    \quad
    \nu(u) =-(au^3+ bu) +\frac{g_S}{u},
\end{eqnarray}
where $a,b$ and $g_S$  are real coupling constants.
\item
Trigonometric model,
\begin{eqnarray}
    x(u)= \cot u, \quad
    z(u) = \frac{1}{\sin^2 u}, \quad
    \nu(u) = a \sin2u -{b\over{\sin2u}} +g_S \cot u, \label{trifunc}
\end{eqnarray}
where $a$, $b$  and $g_S$ are real coupling constants.
\end{enumerate}
The functions $x$ and $\nu$ correspond to  those appearing in
$\partial W/\partial q$.

\subsection{$A$ type models}

The Lax pair is again a pair of $2r \times 2r$ ($r-1$ is the rank) matrices,
with the decomposition
\begin{eqnarray*}
    L = P + X, \quad M = D + Y,
\end{eqnarray*}
in which $P$ and $D$ are diagonal matrices of the form
\begin{eqnarray*}
    P = \sum_{j=1}^r p_j(E_{j,j} - E_{j+r,j+r}), \quad
    D = \sum_{j=1}^r D_j(E_{j,j} + E_{j+r,j+r})
\end{eqnarray*}
and $X$ and $Y$ are diagonal-free matrices of
the form
\begin{eqnarray}
  X &=& ig\sum_{j\not= k} x(q_j - q_k)E_{j,k}
      + ig\sum_{j\not= k} x(-q_j + q_k)E_{j+r,k+r}
    \nonumber \\
    && \mbox{}
      + 2\sum_j \kappa(q_j)E_{j,j+r}
      + 2\sum_j \kappa(-q_j)E_{j+r,j},
    \\
  Y &=& ig\sum_{j\not= k} y(q_j - q_k)E_{j,k}
      + ig\sum_{j\not= k} y(-q_j + q_k)E_{j+r,k+r}
    \nonumber \\
    && \mbox{}
      + \sum_j \kappa'(q_j)E_{j,j+r}
      + \sum_j \kappa'(-q_j)E_{j+r,j}.
\end{eqnarray}
The diagonal elements of $D$ are given by
\begin{eqnarray}
  D_j = - ig\sum_{k\not= j}^rz(q_j - q_k).
\end{eqnarray}

The rational and trigonometric models correspond to the
following choice of functions:
\begin{enumerate}
\item
Rational model,
\begin{eqnarray}
    x(u) = \frac{1}{u}, \quad
    z(u) = \frac{1}{u^2}, \quad
    \kappa(u) = au^2+bu,
\end{eqnarray}
where $a$ and $b$ are arbitrary real coupling
constants.
\item
Trigonometric model,
\begin{eqnarray}
    x(u) = \frac{1}{\sin u}, \quad
    z(u) = \frac{1}{\sin^2 u}, \quad
    \kappa(u) = a\sin2u,
\end{eqnarray}
where $a$ is the arbitrary real coupling
constant.
\end{enumerate}

\section*{Appendix B: Lower triangularity}
\setcounter{equation}{0}
\renewcommand{\theequation}{B.\arabic{equation}}
Here we show the details of the argument that
Calogero-Moser part of the similarity
transformed Hamiltonian (\ref{eq:simtrBCtri}) is lower triangular in the
basis
(\ref{cosbasis}). The triangularity of the trigonometric Calogero-Moser
model
is proved universally in \cite{kps} by using the Coxeter (Weyl)
invariant basis:
\begin{equation}
   \phi_{\lambda}(q)\equiv\sum_{\mu\in W(\lambda)}e^{2i\mu\cdot q},
\label{phidef}
\end{equation}
 in which \(\lambda\) is a
dominant weight
\begin{equation}
   \lambda=\sum_{j=1}^rm_j\lambda_j,\quad m_j\in{\bf Z}_+,\quad \lambda_j:
\mbox{fundamental weight}
\end{equation}
and \(W({\lambda})\) is the orbit of \(\lambda\)
by the action of the Weyl group:
\begin{equation}
   W({\lambda})=\{\mu\in\Lambda(\Delta)|\quad \mu=g(\lambda),\quad
   \forall g\in G_{\Delta}\}, \quad \Lambda(\Delta):
\mbox{weight lattice of}\ \Delta.
\end{equation}
The above \(\phi_{\lambda}(q)\) is Coxeter invariant. The set of functions
\(\{\phi_{\lambda}\}\)  has an order
\(>\):
\begin{equation}
   |\lambda|^2>|\lambda^\prime|^2\quad
   \Rightarrow\quad \phi_{\lambda}>\phi_{\lambda^\prime}.
\end{equation}
For the $BC_r$ root system, the set of weights $W(\lambda)$ is symmetric:
\begin{equation}
\mu\in W(\lambda) \Leftrightarrow -\mu\in W(\lambda).
\label{evenW}
\end{equation}
Thus the Coxeter invariant basis (\ref{phidef}) for
$BC$ type root system can be rewritten
as
\begin{equation}
 \phi_{\lambda}^\prime(q)\equiv\sum_{\mu\in W(\lambda)}\cos({2\mu\cdot q}).
\label{phidef2}
\end{equation}
All the fundamental weights, except for the
spinor weights of $B_r$ are integral.
That is, so long as  the above dominant weight
 $\lambda$ contains the fundamental
spinor weights in {\em even\/} multiples, all the $\mu\cdot q$ in
(\ref{phidef}) have the form
\[
\mu\cdot q=\sum_{j=1}^r k_jq_j,\quad k_j\in{\bf Z}_+.
\]
Therefore the basis
(\ref{phidef2}) has the form (\ref{cosbasis}) used in section
\ref{sec.rBCtri}. Moreover, after the application of $\widetilde{H}$
on the basis functions (\ref{cosbasis}), those appearing as lower terms
have also the same property of having {\em even\/} number of
fundamental spinor weights and can be expressed in the same form
(\ref{cosbasis}). This completes the proof of the lower triangularity of
the Hamiltonian (\ref{eq:simtrBCtri}) in the space (\ref{cosbasis}).


\begin{thebibliography}{99}

\bibitem{bib:ad77}
M.~Adler,
``Some finite-dimensional integrable systems and
their scattering behavior'',
Commun. Math. Phys. {\bf 55} (1977) 195-230.

\bibitem{bib:in83}
V.I.~Inozemtsev,
``On the motion of classical integrable systems
of interacting particles in an external field'',
Phys. Lett. {\bf A98} (1983) 316-318.
 %
\bibitem{bib:in84}
V.I.~Inozemtsev,
``New completely integrable multiparticle
dynamical systems'',
Physica Scripta {\bf 29} (1984) 518-520.

\bibitem{bib:le-wo84}
D.~Levi and S.~Wojciechowsky,
``On the Olshanetsky-Perelomov many-body systems
in an external field'',
Phys. Lett. {\bf A103} (1984) 11-14.



\bibitem{Ino0}
V.I.~Inozemtsev and D.\,V.~Meshcheryakov, ``Extension of the
class of integrable dynamical systems connected
with semisimple Lie algebras",
Lett. Math. Phys. {\bf 9} (1985) 13-18;
V.I.~Inozemtsev, ``Lax representation with
spectral  parameter on a torus for integrable particle
systems'', Lett. Math. Phys. {\bf 17} (1989) 11-17.



\bibitem{Cal}  F.\, Calogero,
 ``Solution of the one-dimensional
\(N\)-body problem with quadratic and/or inversely quadratic pair
potentials",
J. Math. Phys. {\bf 12} (1971)  419-436.
\bibitem{Sut}
B.\, Sutherland,
``Exact results for a quantum many-body problem in
one-dimension. II'',
Phys. Rev. {\bf A5} (1972) 1372-1376. %
\bibitem{CalMo}
J.~Moser,
``Three integrable Hamiltonian systems connected with
isospectral deformations'',
Adv. Math. {\bf 16} (1975) 197-220 ;\ %
``Integrable systems of non-linear evolution equations",
in {\it Dynamical Systems, Theory and Applications\/};\
J.~Moser, ed., Lecture Notes in Physics {\bf 38} (1975),
Springer-Verlag.


\bibitem{OP1} M.\,A.~Olshanetsky and A.\,M.~Perelomov,
``Completely integrable Hamiltonian systems connected with
 semisimple Lie algebras",
 Inventions Math. {\bf 37} (1976) 93-108 ; 
``Classical integrable finite-dimensional systems related to Lie
 algebras'',
Phys. Rep.  {\bf C71} (1981) 314-400.%

\bibitem{OPq}
M.\,A.~Olshanetsky and A.\,M.~Perelomov,
 ``Quantum integrable systems related to Lie algebras",
Phys. Rep. {\bf C94} (1983) 313-404.


\bibitem{bcs2}  A.\,J.~Bordner, E.~Corrigan and R.~Sasaki,
``Generalized Calogero-Moser models and  universal Lax pair operators'',
 Prog. Theor. Phys. {\bf 102} (1999) 499-529,
{\tt  hep-th/9905011}.

\bibitem{DHoker_Phong}
E.~D'Hoker and D.\,H.~Phong,
``Calogero-Moser
Lax pairs with spectral parameter for general Lie algebras'',
Nucl. Phys. {\bf B530} (1998) 537-610, {\tt hep-th/9804124}.

\bibitem{bcs1}
A.\,J.~Bordner, E.~Corrigan and R.~Sasaki,
``Calogero-Moser models I: a new formulation'',
Prog. Theor. Phys. {\bf 100} (1998) 1107-1129,  {\tt hep-th/9805106}.

\bibitem{bst}
A.\,J.~Bordner,   R.~Sasaki and K.~Takasaki,
``Calogero-Moser models II:
symmetries and foldings'',
Prog. Theor. Phys. {\bf 101} (1999) 487-518, {\tt hep-th/9809068}.

\bibitem{sasya}
R.~Sasaki and I.~Yamanaka,
``Virasoro Algebra, Vertex Operators, Quantum Sine-Gordon  and
    Solvable   Quantum   Field  Theories",
    { Adv. Stud. in  Pure
    Math.}  {\bf 16} (1987) 271-296.

\bibitem{kps}
S.\, P.~Khastgir, A.\, J.~Pocklington and R.~Sasaki,
``Quantum Calogero-Moser Models: Integrability for all Root Systems'',
J.\ Phys. {\bf A33} (2000) 9033-9064,
{\tt hep-th/0005277}.

\bibitem{HeOp}
 G. J. Heckman and E. M. Opdam, ``Root
systems and hypergeometric functions I'', Comp. Math. {\bf
64} (1987), 329--352;
G. J. Heckman, ``Root systems and
hypergeometric functions II'', Comp. Math. {\bf 64}
(1987), 353--373;
E. M. Opdam, `` Root systems and
hypergeometric functions III'', Comp. Math. {\bf 67}
(1988), 21--49;
``Root systems and
hypergeometric functions IV'', Comp.  Math. {\bf 67}
(1988), 191--209.


\bibitem{bms}
A.J.~Bordner, N.S.~Manton and R.~Sasaki,
Calogero-Moser models V: supersymmetry and
quantum Lax pair,
Prog. Theor. Phys. {\bf 103} (2000) 463-487,
{\tt hep-th/9910033}.

\bibitem{turush}
A.~Turbiner and A.\,G.~Ushveridze,
``Spectral singularities and  quasi-exactly solvable quantal problem",
Phys. Lett. {\bf A126} (1987) 181-183.

\bibitem{ais}
A.~A.~Andrianov, M.~V.~Ioffe and V.~P.~Spiridonov,
``Higher derivative supersymmetry and the Witten index,''
Phys.\ Lett.\  {\bf A174} (1993) 273-279,
{\tt hep-th/9303005}.

\bibitem{plyu}
S.~M.~Klishevich and M.~S.~Plyushchay,
``Nonlinear supersymmetry on the plane in magnetic field and
 quasi-exactly solvable systems'',
hep-th/0105135;
``Supersymmetry of parafermions'',
Mod.\ Phys.\ Lett.\  {\bf A14} (1999) 2739
-2752, {\tt hep-th/9905149}.


\bibitem{aoy}
H.~Aoyama, M.~Sato and T.~Tanaka,
``$\mathcal{N}$-fold supersymmetry for a periodic potential'',
Phys.\ Lett.\ {\bf B498} (2001) 117-122,
{\tt quant-ph/0011009};
``General forms of an $\mathcal{N}$-fold supersymmetry family'',
 Phys.\ Lett.\ {\bf B503} (2001) 423-429,
{\tt quant-ph/0012065};
H.~Aoyama, N.~Nakayama, M.~Sato and T.~Tanaka,
``sl(2) construction of type-A N-fold supersymmetry,''
hep-th/0107048.




\bibitem{muw}
D.~Mayer, A.~Ushveridze and Z.~Walczak,
``On time-dependent quasi-exactly solvable models'',
Mod.\ Phys.\ Lett.\  {\bf A15} (2000) 1243-1252,
{\tt hep-th/9806192}.

\bibitem{ddt}
P.~Dorey, C.~Dunning and R.~Tateo,
``Spectral equivalences from Bethe ansatz equations,''
J.\ Phys.\  {\bf A34} (2001) 5679-5704,
{\tt hep-th/0103051}.

\bibitem{jsuz}
 J.~Suzuki
`` Functional relations in Stokes multipliers - Fun with
$x^6 + \alpha x^2$ potential-'', J. Stat. Phys. {\bf 102} (2001)
1029-1047, {\tt quant-ph/0003066}.

\bibitem{bender}
C.~M.~Bender and G.~V.~Dunne,
``Quasi-Exactly Solvable Systems and Orthogonal Polynomials,''
J.\ Math.\ Phys.\  {\bf 37} (1996) 6-11,
{\tt hep-th/9511138}.

\bibitem{razavy}
M.~Razavy, ``An exactly soluble Schr\"odinger equation with a bistable
potential'',
Amer. J. Phys. {\bf 48} (1980) 285-288.

\bibitem{stie}
T.\,J.~Stieltjies, ``Sur quelques th\'eor\`emes d'alg\`ebre",
Compte Rendus {\bf
100} (1885) 439-440; ``Sur les polyn\^omes de Jacobi", ibid. 620-622;
G.~Szeg\"o,
``Orthogonal Polynomials", Amer. Math. Soc. (1939),
\S6.7;\ F.~Calogero,
``Equilibrium configuration of the one-dimensional $n$-body
problem with quadratic and inverse quadratic pair potentials",
Lett. Nuovo Cim. {\bf 20} (1977) 251-253.


\bibitem{hs1}
 X.~Hou and  M.~Shifman, ``A quasiexactly solvable N body problem
 with the SL(N+1)
algebraic structure",
Int. J. Mod. Phys. {\bf A14} (1999) 2993-3004,
{\tt  hep-th/9812157}.

\bibitem{turb}
A.\,V.~Turbiner, ``Quasi-exactly-soluble problems and sl(2,R) algebra",
Comm.
Math. Phys. {\bf 118} (1988) 467-474.


\bibitem{Oshima}
T.\, Oshima and H.\, Sekiguchi, ``Commuting families of differential
operators invariant under the action of a Weyl group",
J. Math. Sci. Univ. Tokyo
{\bf 2}  (1995) 1-75.



\bibitem{FinkelAN}
F.~Finkel, D.~Gomez-Ullate, A.~Gonzalez-Lopez, M.~A.~Rodriguez and
 R.~Zhdanov,
``A(N)-type Dunkl operators and new spin Calogero-Sutherland models",
{\tt hep-th/0102039};
``New spin Calogero-Sutherland models related to $B_N$-type Dunkl
 operators'',
{\tt hep-th/0103190}.


\bibitem{khare}
A.~Khare, ``A QES band-structure problem in one dimension",
{\tt quant-ph/0105030}.

\end{thebibliography}
\end{document}